\documentclass[10pt]{article}
\usepackage{authblk}
\usepackage[bookmarksnumbered, colorlinks, plainpages]{hyperref}
\usepackage{amsmath, amsthm, amscd, amsfonts, amssymb, graphicx, color, booktabs}
\numberwithin{equation}{section}
\definecolor{email}{rgb}{0.00,0.00,0.84}

\newcommand{\fbinv}{\mathrm{fb}^{-1}}

\newcommand{\GeV}{\mathrm{{GeV}}}
\newcommand{\TeV}{\mathrm{{TeV}}}

\newcommand{\Pp}{\ensuremath{{p}}}
\newcommand{\PH}{\ensuremath{{H}}}
\newcommand{\Hgg}{\ensuremath{H \rightarrow \gamma\gamma}}

\newcommand{\PZ}{\ensuremath{{Z}}}

\newcommand{\cPqt}{\ensuremath{{t}}}

\newcommand{\mH}{\ensuremath{m_{H}}}

\newcommand{\PGg}{\ensuremath{{\gamma}}}

\newcommand{\br}{\mathcal{B}}

\newcommand{\PGt}{\ensuremath{{\tau}}}
\newcommand{\ditau}{\ensuremath{\PGt\PGt}}
\newcommand{\PQb}{\ensuremath{{b}}}
\newcommand{\Pphi}{\ensuremath{{\phi}}}
\newcommand{\Pg}{\ensuremath{{g}}}
\newcommand{\ggPhi}{\ensuremath{\Pg\Pg\phi}}
\newcommand{\mphi}{\ensuremath{m_{\phi}}}

\newcommand{\Pe}{\ensuremath{{e}}}
\newcommand{\Zphi}{\ensuremath{{\PZ}{\phi}}}

\newcommand{\ttbar}{\ensuremath{\cPqt\bar{\cPqt}}}
\newcommand{\tta}{\ensuremath{{\ttbar}a}}

\newcommand{\PHpm}{\ensuremath{{H}^{\pm}}}
\newcommand{\mHpm}{\ensuremath{m_{\PHpm}}}

\newcommand{\PGm}{\ensuremath{{\mu}}}

\newcommand{\PX}{\ensuremath{{X}}}
\newcommand{\sppXem}{\ensuremath{{\sigma(\Pp\Pp \to \PX \to \Pe \PGm)}}}
\newcommand{\mx}{\ensuremath{{m_{\PX}}}}

\newcommand{\cPb}{\ensuremath{{b}}}
\newcommand{\bbbar}{\ensuremath{\cPb\bar{\cPb}}}
\newcommand{\bbbb}{\ensuremath{\bbbar\bbbar}}

\newcommand{\PY}{\ensuremath{{Y}}}
\newcommand{\HH}{\ensuremath{{\PH\PH}}}

\newcommand{\mX}{\ensuremath{{m_{\PX}}}}
\newcommand{\mY}{\ensuremath{{m_{\PY}}}}
\newcommand{\bbgg}{\ensuremath{\PGg\PGg\bbbar}}
\newcommand{\ppYHbbgg}{\ensuremath{\Pp\Pp \to \PX \to \PH\PY \to \bbgg}}

\newcommand\blfootnote[1]{%
  \begingroup
  \renewcommand\thefootnote{}\footnote{#1}%
  \addtocounter{footnote}{-1}%
  \endgroup
}

\def\institute{Institute of High Energy Physics, Chinese Academy of Sciences}
\def\authemail{\footnote{Contact: taojq@ihep.ac.cn}}

\begin{document}

\setcounter{page}{1}

\title{\large \bf 12th Workshop on the CKM Unitarity Triangle\\ Santiago de Compostela, 18-22 September 2023 \\ \vspace{0.3cm}
\LARGE Extra Higgs boson searches at the LHC}

\author{Junquan Tao\authemail $ $
 \  on behalf of the ATLAS and CMS Collaborations}
\affil{\institute}

\maketitle

\begin{abstract}
\noindent
Many searches for additional Higgs bosons, which are predicted by a lot of interesting models beyond the standard model, have been performed at the LHC. Some selected latest results of the searches for extra Higgs bosons at the LHC are presented. These additional Higgs bosons could be produced either directly from the parton interactions or from the decays of the observed standard model Higgs boson or a new heavier resonance. The searches used the data from proton-proton collisions delivered by the LHC at a centre-of-mass energy of $\sqrt{s}=13~\TeV$ and recorded with the ATLAS and CMS detectors. No direct evidence of new physics has been observed yet. Several mild excesses were observed in some final states. More data is needed to conclude on the nature of these excesses.
\end{abstract} \maketitle

\blfootnote{Copyright 2024 CERN for the benefit of the ATLAS and CMS Collaborations. Reproduction of this article or parts of it is allowed as specified in the CC-BY-4.0 license.}

\section{Introduction}

\noindent 
The standard model (SM)
~\cite{Glashow:1961tr,Weinberg:1967tq,Salam:1968rm}
of particle physics
has been very successful in explaining high-energy experimental data.
In 2012, a new particle  with a mass of
about 125 $\GeV$
was discovered~\cite{Aad:2012tfa,Chatrchyan:2012xdj,Chatrchyan:2013lba} by both the ATLAS and CMS experiments. 
The properties of the new particle are so far compatible with those of the
SM Higgs boson~\cite{CMS:2022dwd,ATLAS:2022vkf},
the quantum of the scalar field postulated by the
Brout--Englert--Higgs mechanism~\cite{Englert:1964et,Higgs:1964pj,Guralnik:1964eu}.
Many models for physics beyond the SM (BSM), such as the generalized
models containing two Higgs doublets~\cite{Celis:2013rcs,Cacciapaglia:2016tlr},
the next-to-minimal supersymmetric model~\cite{Ellwanger:2009dp,Fan:2013gjf,Tao:2018zkx}, 
and Georgi-Machacek model~\cite{Georgi:1985nv,Wang:2022okq}, 
can provide a Higgs boson that is compatible with the observed 125 $\GeV$ boson, and 
additional Higgs bosons with rich and interesting phenomenologies.
So, the discovery of additional scalars including neutral and charged Higgs bosons
would be an unequivocal sign of new physics.

Some latest and selected
results of the 
extra Higgs boson searches at the LHC, are presented.
This paper will mainly focus on the results in which mild excesses with respect to the SM background expectation were observed.
The analyses included in this paper are a biased selection, according to the author's personal favourites.
The data sets utilized are the proton-proton ($\Pp\Pp$) collisions at a center-of-mass energy ($\sqrt{s}$) of 13 $\TeV$,
collected with the ATLAS detector~\cite{ATLAS:2008xda} and CMS
detector~\cite{CMS:2008xjf} at the LHC during the Run 2 data-taking period (2015--2018).

\section{Direct searches for extra Higgs boson}
\begin{figure}[htbp]
\centering
\includegraphics[width=0.418\textwidth]{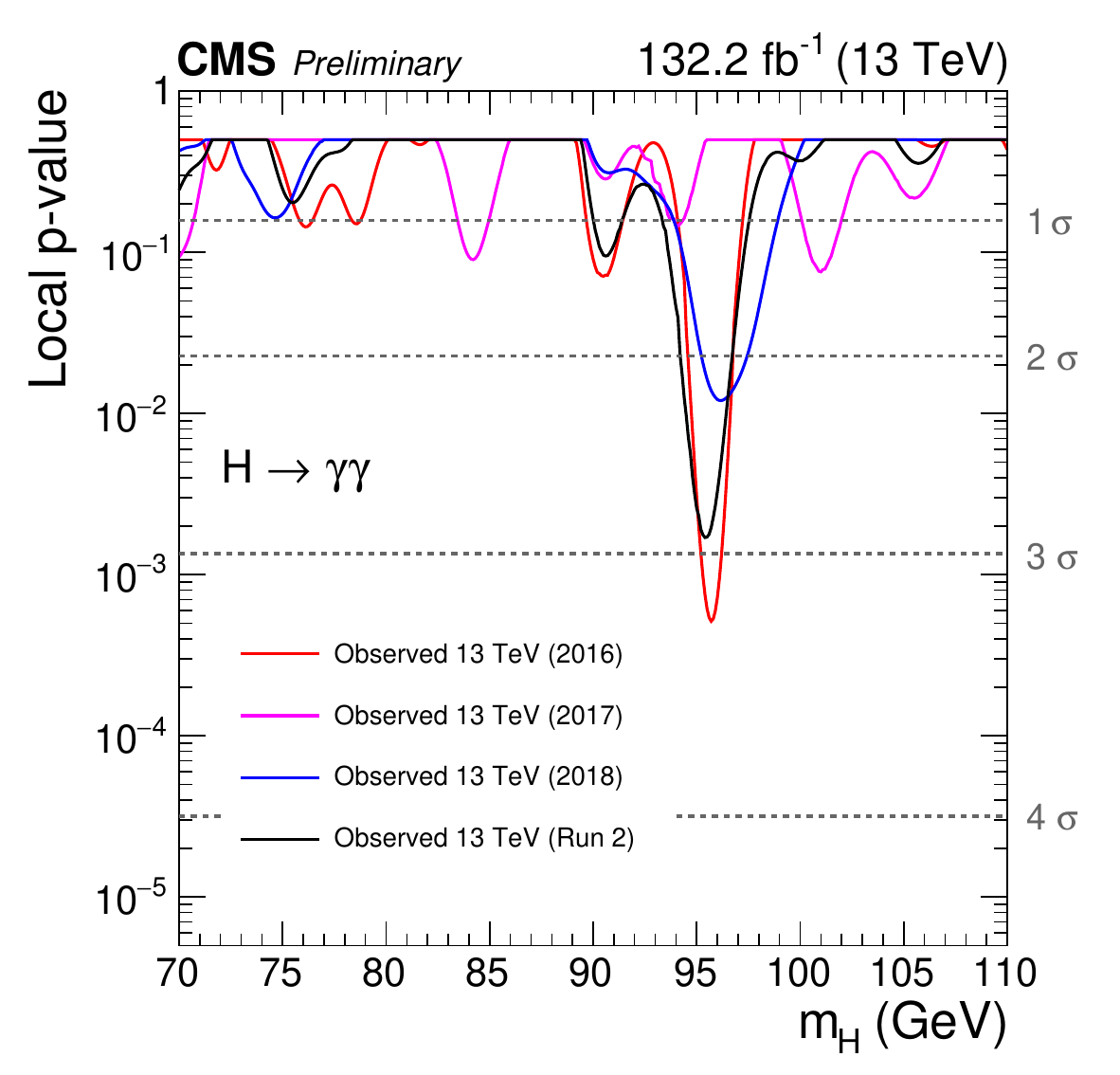}  
\includegraphics[width=0.562\textwidth]{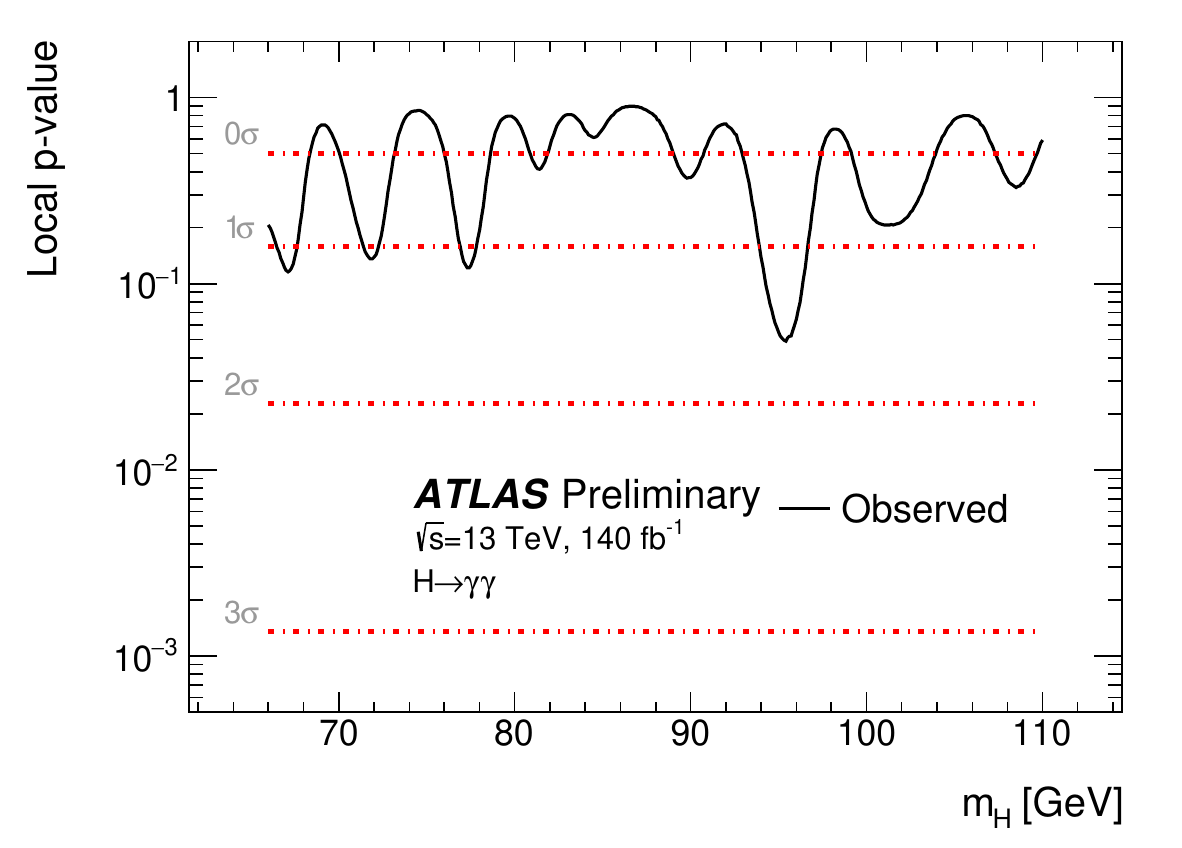} \\ 
\includegraphics[width=0.418\textwidth]{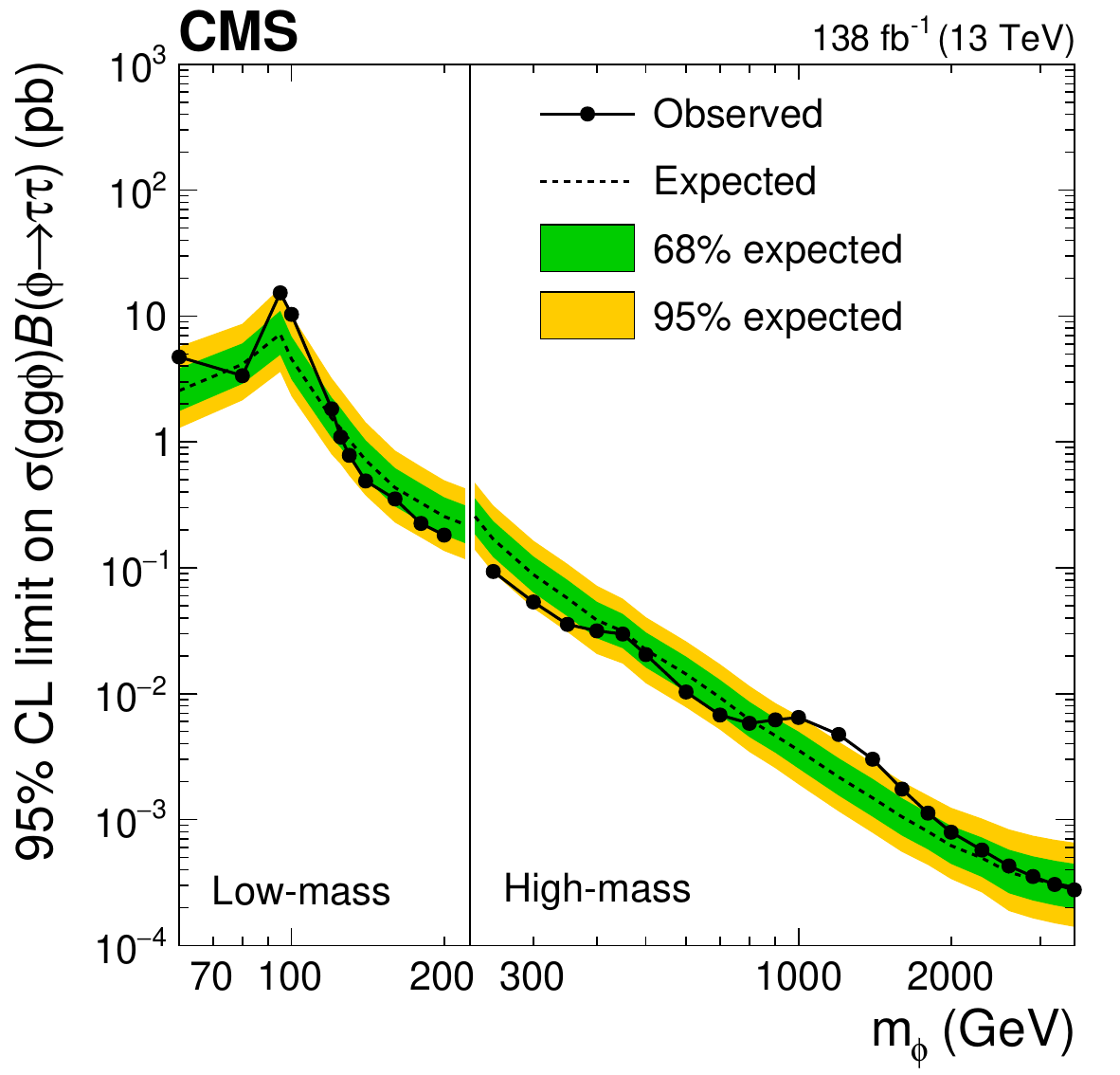}  
\caption{ (color online)
The observed local $p$-values for an additional SM-like
Higgs boson decaying into two photons as a function of $\mH$ reported by CMS (upper left)~\cite{CMS:2023yay} and ATLAS (upper right)~\cite{ATLAS:2023jzc},
based on the LHC Run 2 data set.
(Lower) Expected and observed 95\% CL upper limits on the product of the cross
    sections and branching fraction for the decay into $\PGt$ leptons for
    $\ggPhi$ production in a mass range of $60\leq\mphi\leq
    3500\GeV$, reported by CMS with the full Run 2 data set~\cite{CMS:2022goy}.
} \label{fig:LM_Higgs}
\end{figure}

\noindent
The $\Hgg$ decay channel provides a clean final-state topology
that allows the mass of a Higgs boson to be reconstructed with high precision.
A CMS search~\cite{CMS:2018cyk} performed at center-of-mass energies of 8 and 13 $\TeV$ (2016) 
reported an excess with respect to the
SM background prediction,
maximal for a mass hypothesis of 95.3 $\GeV$, with a local (global) significance of 2.8\,(1.3) standard deviations ($\sigma$).
Based on the data collected at a center-of-mass energy of 13 TeV during the LHC Run 2 with the corresponding integrated luminosity $132~\mathrm{fb}^{-1}$,
CMS reported the updated results of a search for a standard model-like Higgs boson decaying into two photons in the mass range between 70 and 110$~\mathrm{GeV}$~\cite{CMS:2023yay}.
An excess with respect to the standard model background prediction, which is maximal for a mass hypothesis of 95.4 $\mathrm{GeV}$ with a local (global) significance of 2.9$\sigma$ (1.3$\sigma$), is observed, as shown in Figure~\ref{fig:LM_Higgs} (upper left).
Using $140~\mathrm{fb}^{-1}$ of pp collisions,
ATLAS also released its search results on light, spin-0 bosons decaying to two photons in the 66 to 110$~\mathrm{GeV}$ mass range~\cite{ATLAS:2023jzc}.
In the model-dependent search, the largest deviation is observed for a mass of 95.4 $\mathrm{GeV}$, corresponding to a local significance of 1.7$\sigma$,
as shown in Figure~\ref{fig:LM_Higgs} (upper right).

The $\ditau$ final state has a leading role in the searches for additional heavy neutral Higgs bosons in the context of the minimal supersymmetric SM.
CMS performed the search for a new boson $\Pphi$ via gluon fusion ($\ggPhi$) or in association with $\PQb$ quarks, in $\ditau$ final states
using a data set corresponding to 138 $\fbinv$ collected during LHC Run 2~\cite{CMS:2022goy}.
The data reveal two excesses for $\ggPhi$ production with local $p$-values equivalent to about 3$\sigma$ at $\mphi=0.1$ and 1.2 $\TeV$,
as shown in Figure~\ref{fig:LM_Higgs} (lower).

\begin{figure}[htbp]
\centering
\includegraphics[width=0.41\textwidth]{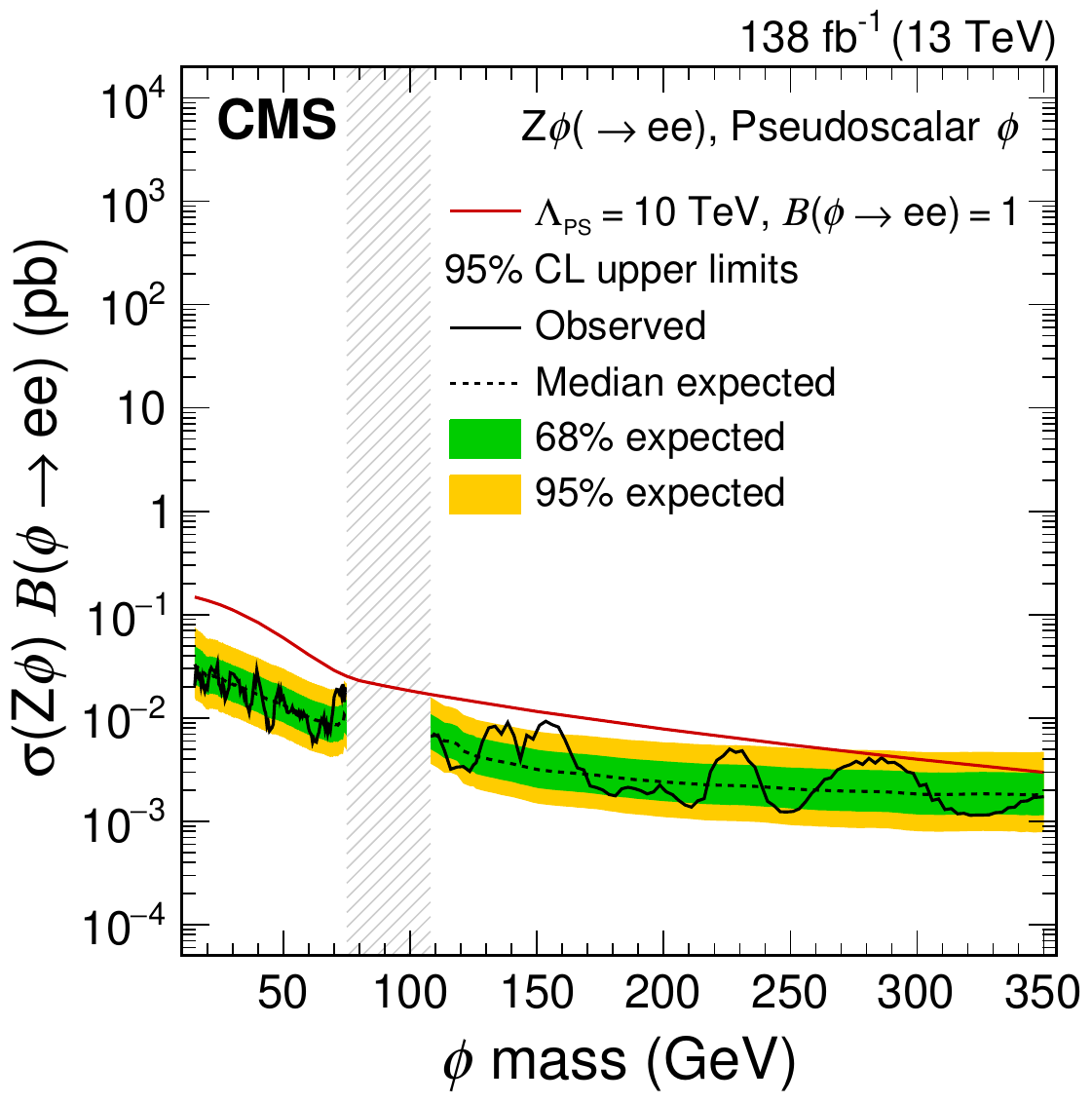}
\includegraphics[width=0.57\textwidth]{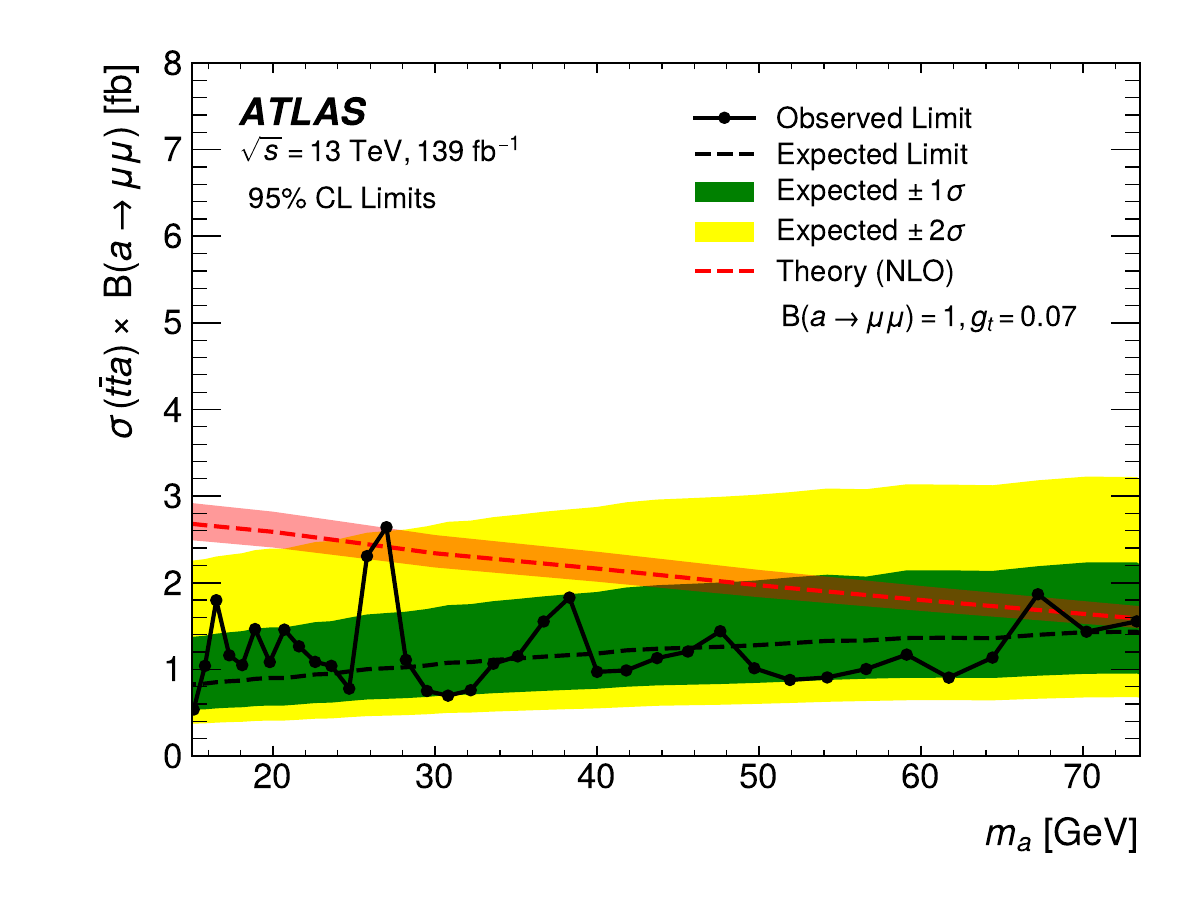}
\caption{ (color online)
(Left) The 95\% confidence level upper limits on the product of the production cross section and branching fraction of the $\Zphi$ signal in the $\Pe\Pe$ decay scenario, reported by CMS with the full Run 2 data set~\cite{CMS:2024ulc}.
(Right) Expected and observed 95\% confidence level (CL) upper limits on the signal cross section shown as a function of $m_{a}$ for $\tta$ production,
reported by ATLAS with the full Run 2 data set~\cite{ATLAS:2023ofo}.
} \label{fig:2}
\end{figure}

CMS also performed the search for BSM spin-0 bosons, $\phi$, that decay into pairs of electrons, muons, or tau leptons,
using a data set corresponding to 138 $\fbinv$ collected during LHC Run 2~\cite{CMS:2024ulc}.
The largest local deviation is observed in the high mass $\Zphi\to\Pe\Pe$ search, as shown in Figure~\ref{fig:2} (left),
where an excess at a $\phi$ mass of 156 $\GeV$ corresponding to 2.9$\sigma$ is observed.
ATLAS performed the search for a new pseudoscalar
$a$-boson produced in events with a top-quark pair, where the
$a$-boson decays into a pair of muons,
using a data set corresponding to 139 $\fbinv$ collected during LHC Run 2~\cite{ATLAS:2023ofo}.
As shown in Figure~\ref{fig:2} (right), the smallest local p-value is 0.008 at $m_{a}$ = 27 $\GeV$, corresponding to a local significance of about 2.4$\sigma$.

\begin{figure}[htbp]
\centering
\includegraphics[width=0.52\textwidth]{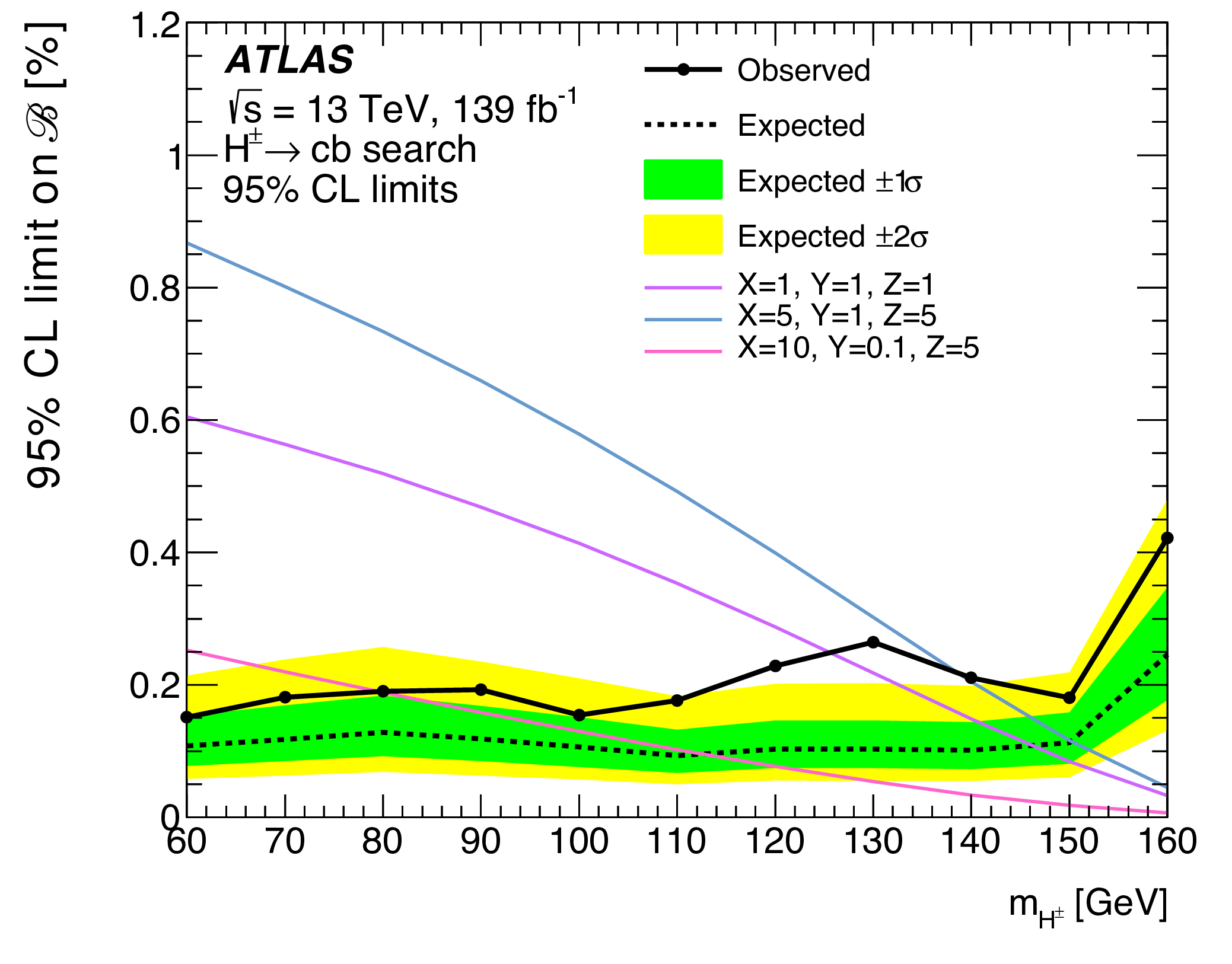}
\includegraphics[width=0.46\textwidth]{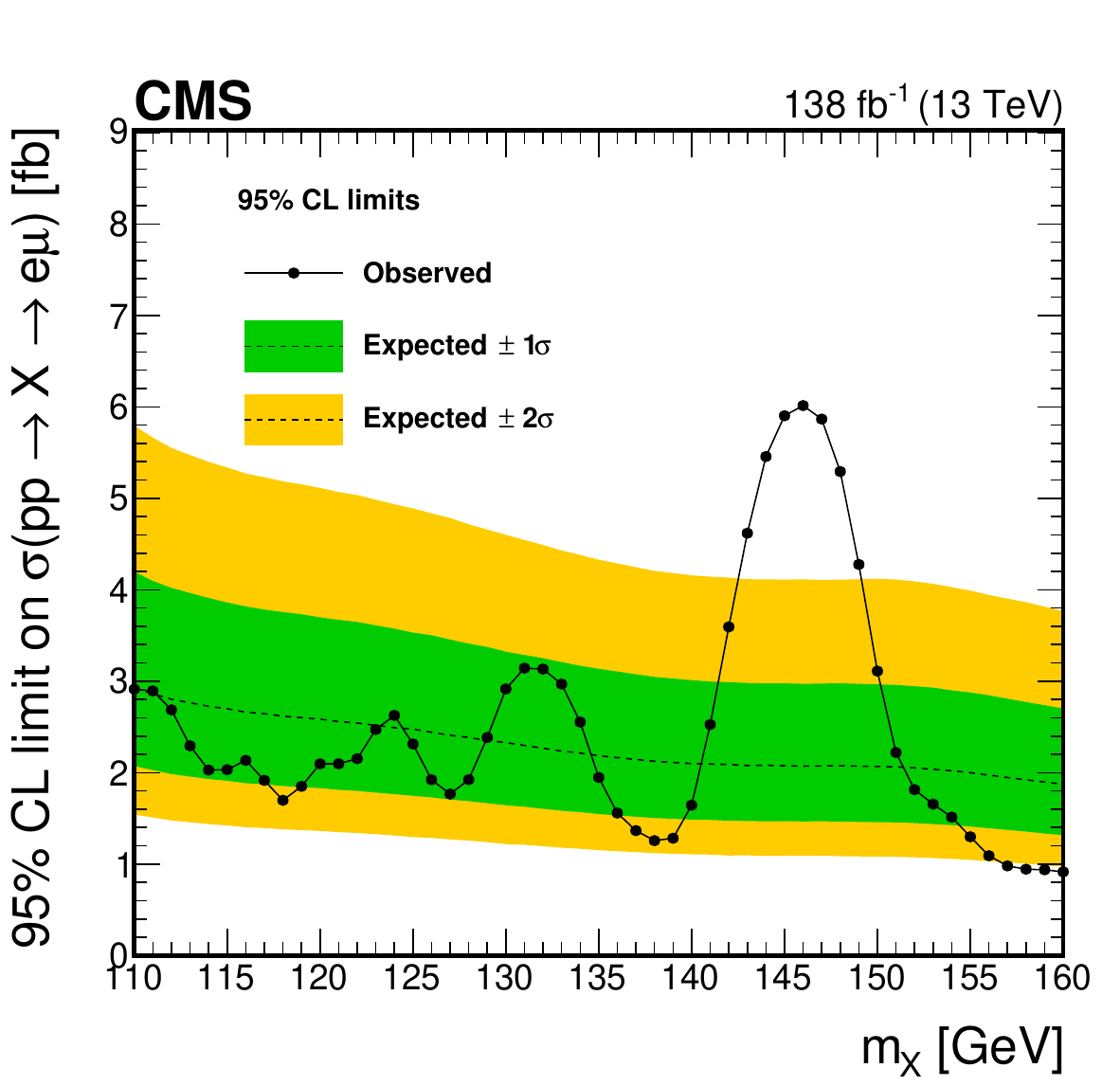}
\caption{ (color online)
(Left) The observed (solid) 95\% CL upper limits on $\br=\br(t\to\PHpm b) \times \br(\PHpm \to cb)$ as a function of $\mHpm$ and the expectation (dashed) under the background-only hypothesis, reported by ATLAS with the full Run 2 data set~\cite{ATLAS:2023bzb}.
(Right) The observed (expected) 95\% CL upper limits on $\sppXem$ as a function of the hypothesized $\mx$ assuming the relative SM-like production cross sections of the
gluon-gluon fusion and vector boson fusion production modes, reported by CMS with the full Run 2 data set~\cite{CMS:2023pte}.
} \label{fig:3}
\end{figure}

ATLAS performed the search for a charged Higgs boson, $\PHpm$, produced in top-quark decays, $t\to\PHpm b$ with $\PHpm$ decays into a bottom and a charm quark,
using a data set corresponding to 139 $\fbinv$ collected during LHC Run 2~\cite{ATLAS:2023bzb}.
Figure~\ref{fig:3} (left) shows the observed (solid) 95\% CL upper limits on $\br=\br(t\to\PHpm b) \times \br(\PHpm \to cb)$ as a function of $\mHpm$ and the expectation (dashed) under the background-only hypothesis.
The largest excess in data has a significance of about 3$\sigma$ for $\mHpm$ = 130 $\GeV$.

CMS performed the search for the lepton-flavor violating decay of the Higgs boson and potential additional Higgs bosons with a mass in the range 110--160 $\GeV$ to an $\Pe^{\pm}\PGm^{\mp}$ pair,
using a data set corresponding to 138 $\fbinv$ collected during LHC Run 2~\cite{CMS:2023pte}.
The largest excess of events over the expected background in the full mass range of the search is observed at an $\Pe^{\pm}\PGm^{\mp}$ invariant mass of approximately {146 $\GeV$} with a local (global) significance of 3.8$\sigma$ (2.8$\sigma$), as shown in Figure~\ref{fig:3} (right).

\section{Searches for light pseudo-scalars in exotic Higgs boson decays}

\noindent
Recent measurements of the Higgs boson's couplings at the LHC do not rule out exotic decays of the Higgs boson to BSM particles.
The exotic decay channels may include the Higgs boson decaying to a pair of light pseudoscalar particles ($H \to aa$)
that subsequently decay to pairs of SM particles.
$H \to aa$ appears in many well-motivated extensions of SM.

\begin{figure}[htbp]
\centering
\includegraphics[width=0.52\textwidth]{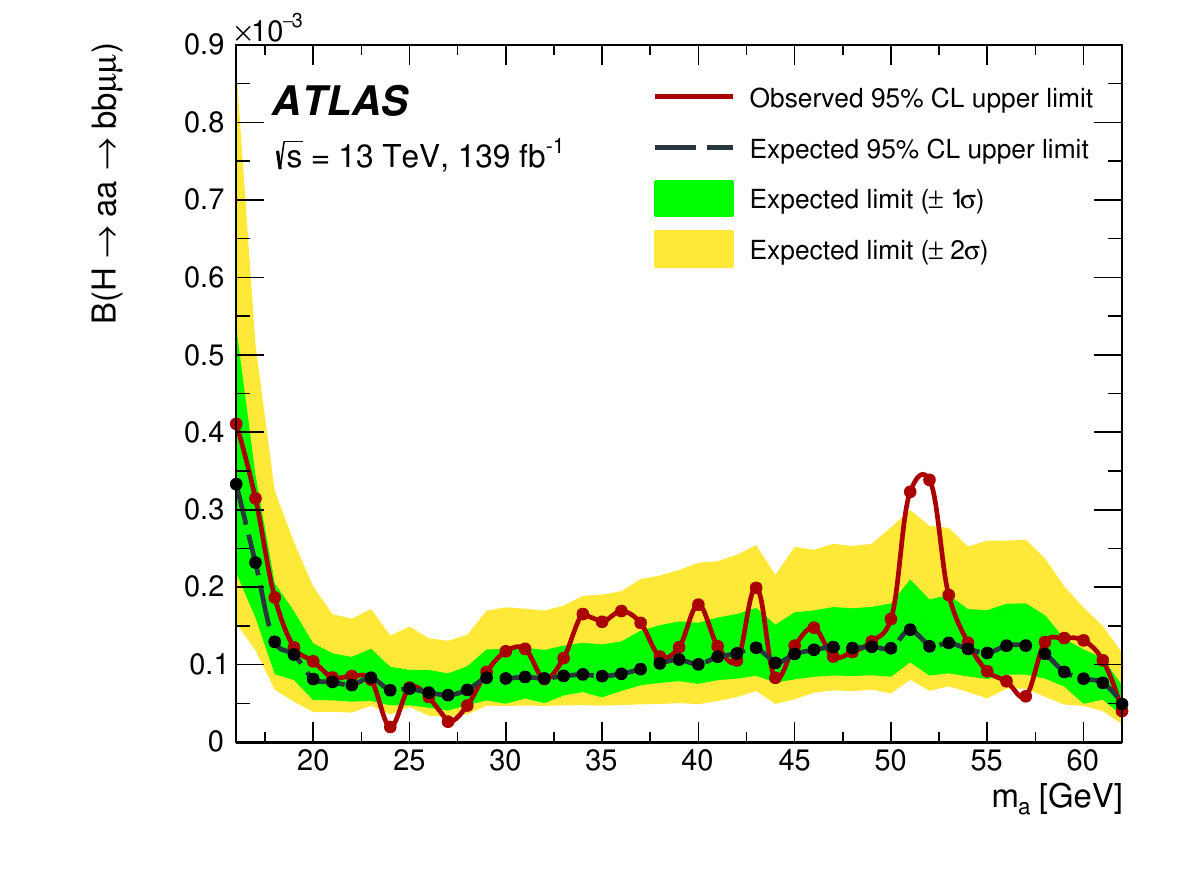}
\includegraphics[width=0.46\textwidth]{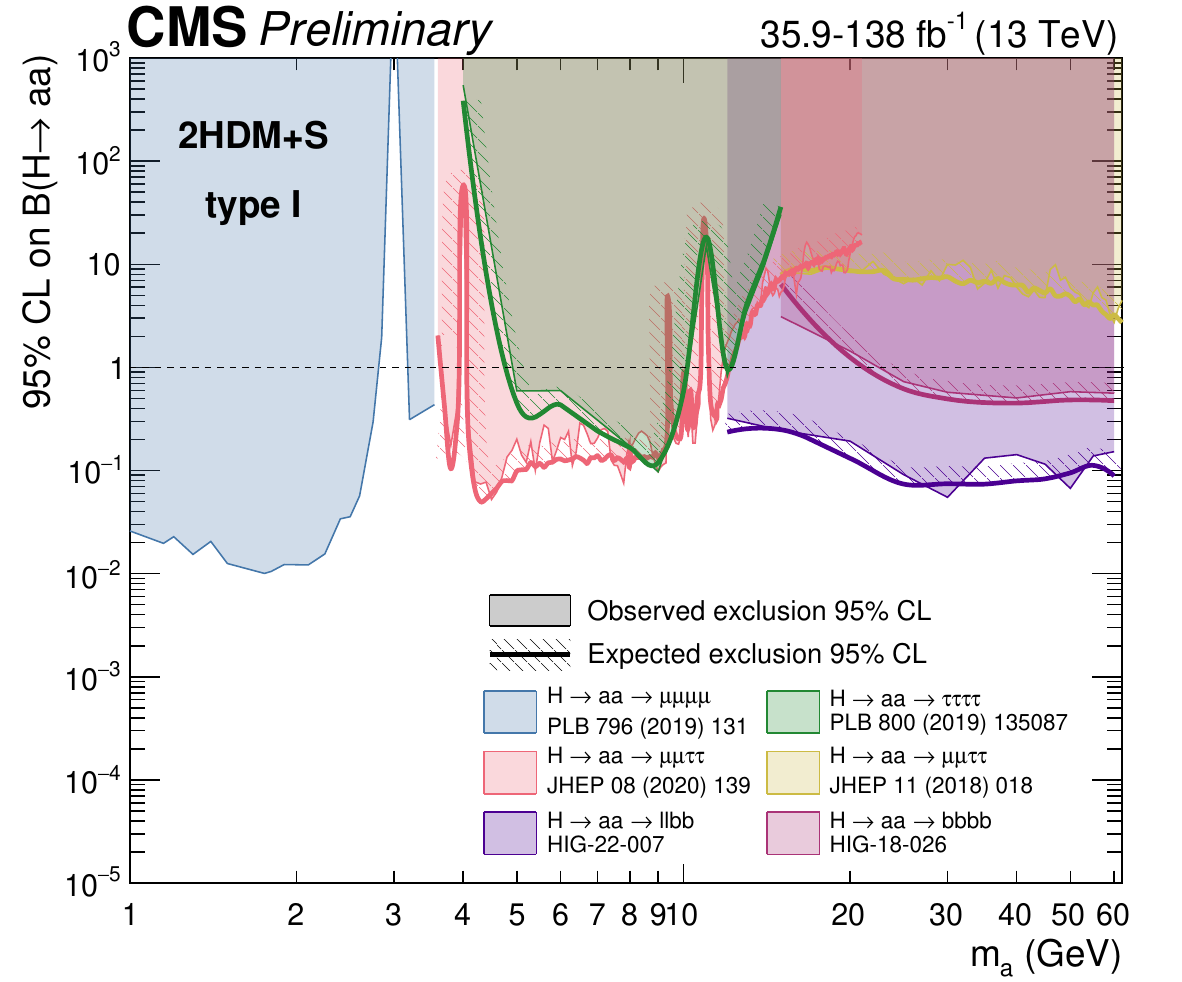}
\caption{ (color online)
(Left) Upper limits on $\br(H \to aa \to bb \mu \mu)$ at 95\% CL, including the BDT selection, as a function of the signal mass hypothesis, reported by ATLAS using full Run 2 data set~\cite{CMS:2023pte}.
(Right) 95\% CL on $\sigma(h)/\sigma_{SM} \times \br(H \to aa)$ in the 2HDM+S type-1 scenario for exotic $H$ decay searches performed with data collected at 13 $\TeV$ center-of-mass energy, summarized from CMS results~\cite{cms:haa}.
} \label{fig:4}
\end{figure}

The ATLAS Collaboration performed a search
for $H \to aa$ decays, where one $a$-boson decays into a $b$-quark pair and the other into a muon pair,
using a pp collision data set corresponding to an integrated luminosity of 139 $\fbinv$~\cite{CMS:2023pte}.
A narrow dimuon resonance is searched for in the invariant mass spectrum between 16 $\GeV$ and 62 $\GeV$.
The largest excess of events above the Standard Model backgrounds is observed at a dimuon invariant mass of 52 GeV and corresponds to a local (global) significance of 3.3 $\sigma$ (1.7$\sigma$).
Upper limits at 95\% CL on $\br(H \to aa \to bb \mu \mu)$ including (excluding) the BDT (boosted decision tree) selection criteria range between $0.2 \times 10^{-4}$ and
$4 \times 10^{-4}$ ($0.5 \times 10^{-4}$  and $5 \times 10^{-4}$), depending on $m_a$, as shown in Figure~\ref{fig:4} (left).

A summarized figure from all the CMS searches for an extension of the Two Higgs Doublet Model (2HDM) with an additional singlet (2HDM+S) in the type-1 scenario using Run 2 data set, is shown in Figure~\ref{fig:4} (right)~\cite{cms:haa}. Upper limits on $\br(H \to aa)$ depend on the 2HDM types and model parameters.

\section{Searches for multi-Higgs boson resonance}

\noindent
Many BSM theories predict the existence of a heavy scalar boson ($X$) decaying into two Higgs bosons ($HH$) or Higgs boson plus additional scalar ($HY$) in the extended Higgs sector. As the Higgs boson can decay in many different ways, both ATLAS and CMS 
performed searches for resonant di-Higgs production
in various decay channels, using the LHC Run 2 $\Pp\Pp$ collision data at $\sqrt{s} = 13\TeV$.

\begin{figure}[htbp]
\centering
\includegraphics[width=0.48\textwidth]{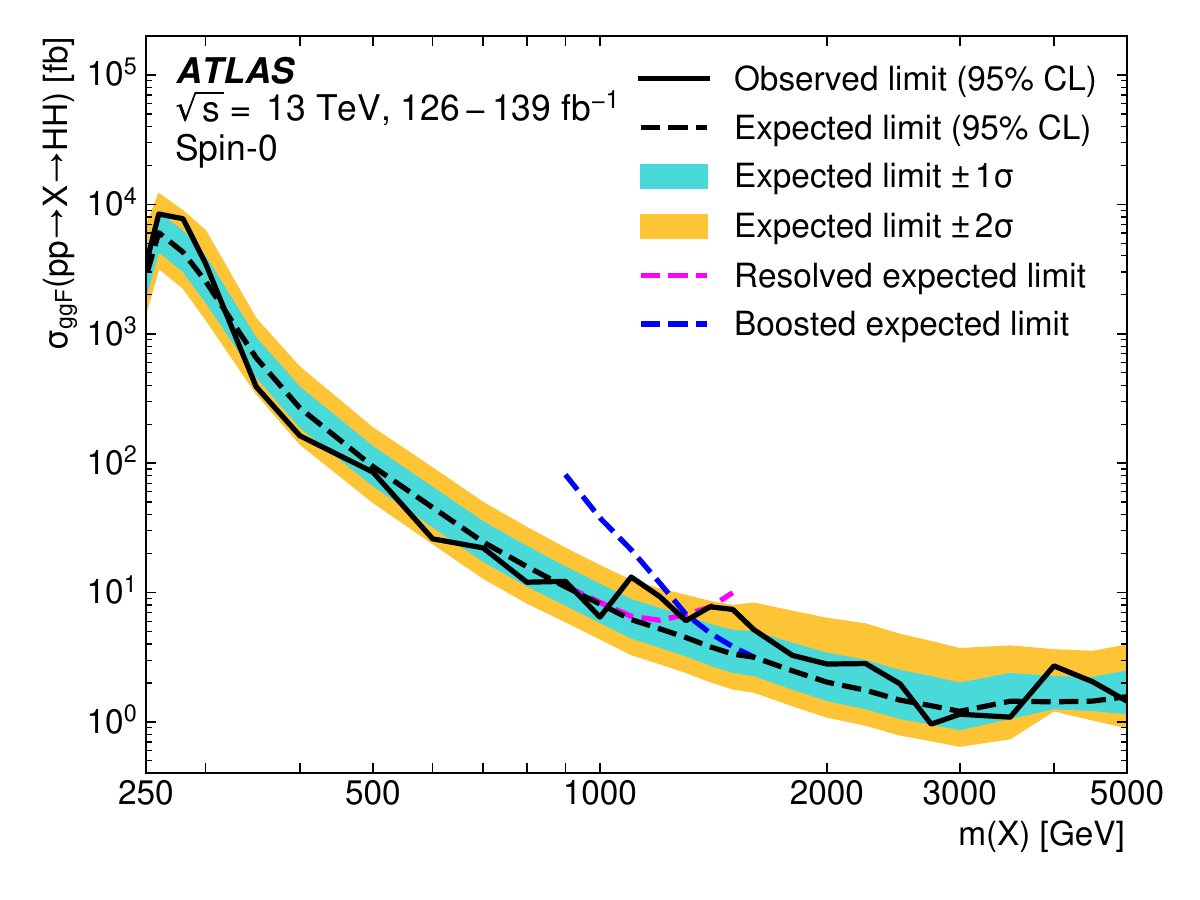}
\includegraphics[width=0.50\textwidth]{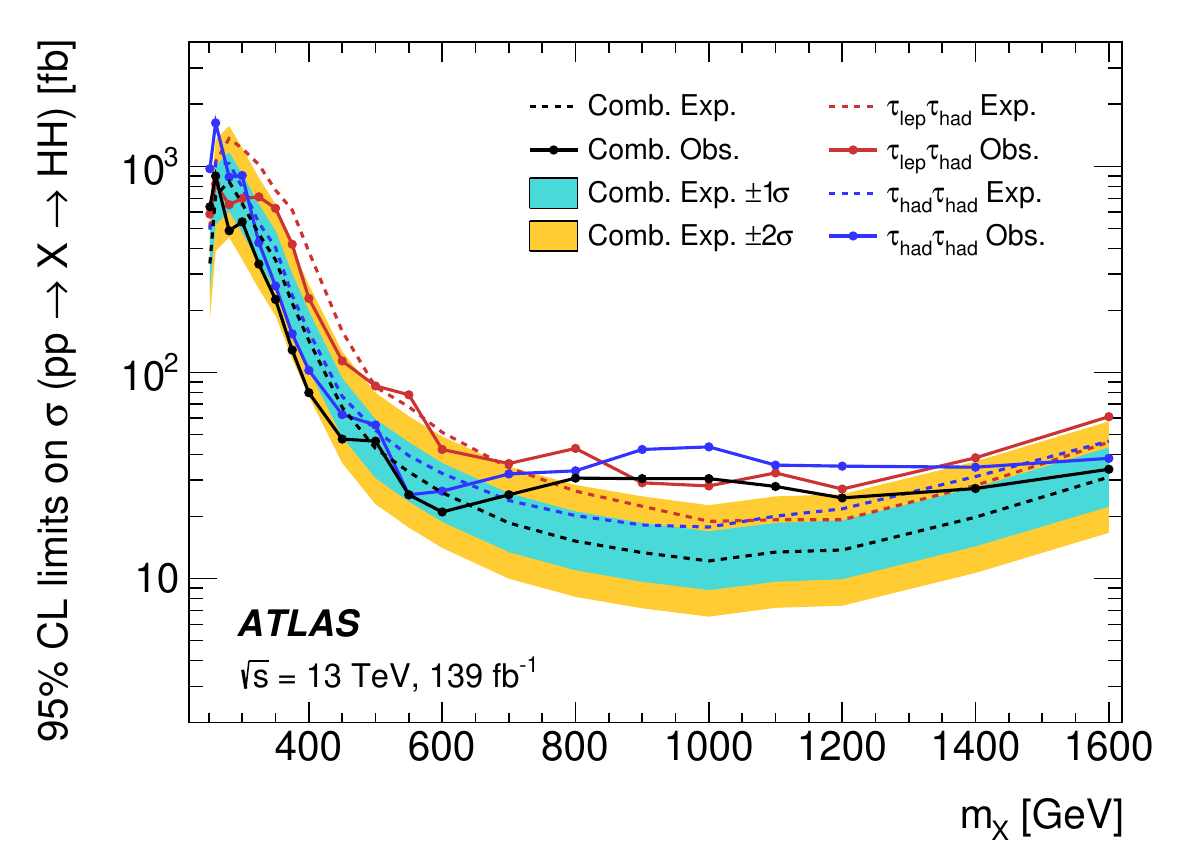}
\caption{ (color online)
(Left) Expected (dashed black lines) and observed (solid black lines) 95\% CL upper limits on the cross-section of resonant HH production in the spin-0 signal models~\cite{ATLAS:2022hwc}.
(Right) Observed (solid lines) and expected (dashed lines) limits at 95\% CL on the cross-section for resonant HH production as a function of the scalar resonance mass $\mx$~\cite{ATLAS:2022xzm}.
All the results are reported by ATLAS using Run 2 data set.
} \label{fig:5}
\end{figure}

The ATLAS Collaboration performed a search
for resonant Higgs boson pair production in the $\bbbb$ final state,
using 126--139 $\fbinv$ of pp collision data at $\sqrt{s} = 13~\TeV$~\cite{ATLAS:2022hwc}.
Spin-0 and spin-2 benchmark signal models are considered, both of which correspond to resonant HH production via gluon-gluon fusion.
The most significant excess is found for a signal mass of 1100 $\GeV$,
with the local significance of 2.3$\sigma$ for the spin-0 signal model, as shown in Figure~\ref{fig:5} (left), and 2.5$\sigma$ for the spin-2 signal model.
ATLAS also performed the search for Higgs boson pair production in events with two $b$-jets and two $\tau$-leptons, using 139 $\fbinv$ pp collision data~\cite{ATLAS:2022xzm}.
Higgs boson pairs produced in the decay of a narrow scalar resonance in the mass range from 251 to 1600 $\GeV$ are targeted.
The largest excess in the resonant search is observed at a resonance mass of 1 $\TeV$, with a local (global) significance of 3.1$\sigma$ (2.0$\sigma$).
Observed (expected) 95\% confidence-level upper limits are set on the resonant Higgs boson pair-production cross-section at between 21 and 900 $fb$ (12 and 840 $fb$),
depending on the mass of the narrow scalar resonance, as shown in Figure~\ref{fig:5} (right).

\begin{figure}[htbp]
\centering
\includegraphics[width=0.48\textwidth]{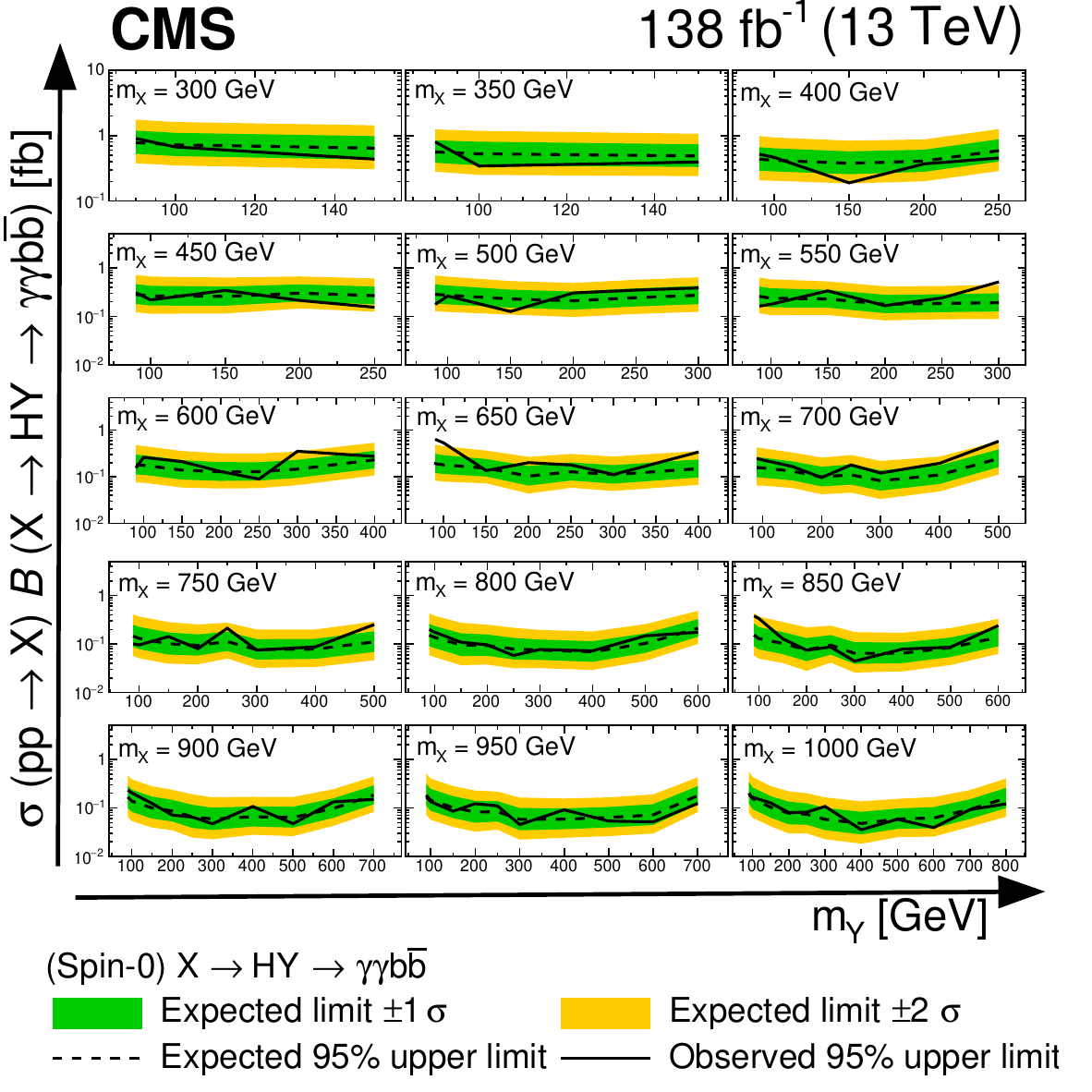}
\includegraphics[width=0.50\textwidth]{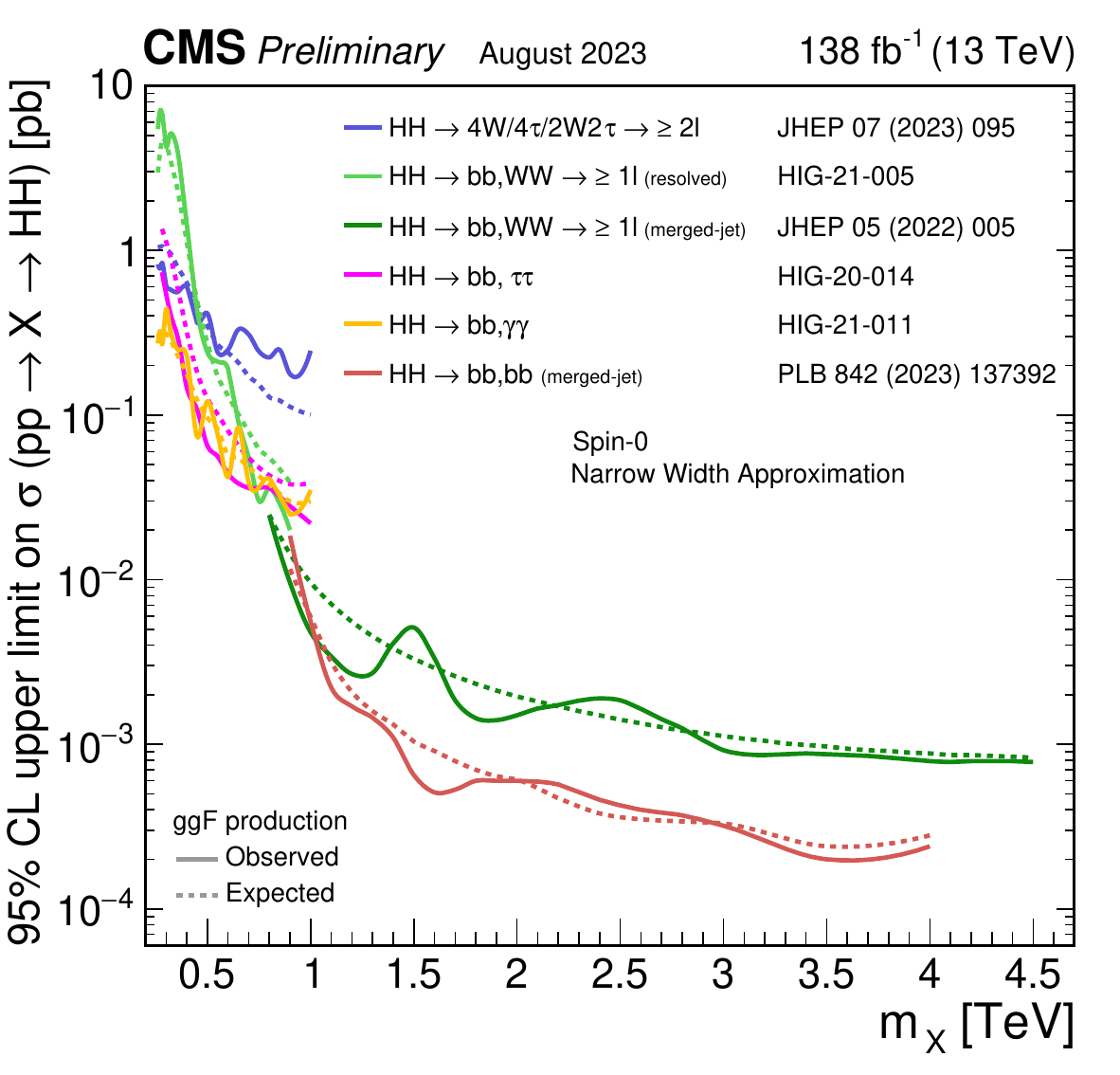}
\caption{ (color online)
(Left) Expected (dashed black lines) and observed (solid black lines) 95\% CL exclusion limit on production cross section for $\ppYHbbgg$ signal~\cite{CMS:2023boe}.
The middle plot in the 3rd row shows the largest excess observed for $\mX$ = 650 $\GeV$ and $\mY$ = 90 $\GeV$.
(Right) Observed and expected 95\% CL upper limits on the product of the production cross section times $\br(HH)$ of a spin-0 resonance ($X$) versus its mass obtained by different analyses~\cite{cms:hh}. All the results are reported by CMS using full Run 2 data set.
} \label{fig:6}
\end{figure}

CMS performed a search for a new boson $\PX$, decaying into either a pair of Higgs bosons $\HH$ of mass 125 $\GeV$ or an $\PH$ and a new spin-0 boson $\PY$, with one $\PH$ subsequently decaying to a pair of photons and the second $\PH$ or $\PY$ to a pair of bottom quarks,
using 138 $\fbinv$  of pp collision~\cite{CMS:2023boe}.
For a spin-0 $\PX$ hypothesis, the 95\% confidence level upper limit on the product of its production cross section and decay branching fraction is 
found to be within 0.90--0.04 $fb$, depending on the masses of $\PX$ and $\PY$ as shown in Figure~\ref{fig:6} (left). The largest deviation from the background-only hypothesis with a local (global) significance of 3.8$\sigma$ (below 2.8$\sigma$) is observed for $\PX$ and $\PY$ masses of 650 and 90 $\GeV$, respectively.
Figure~\ref{fig:6} (right) shows the observed and expected 95\% CL upper limits on the product of the production cross section times $\br(HH)$ of a spin-0 resonance ($X$) versus its mass obtained by different analyses performed by CMS~\cite{cms:hh}.

\section{Summary}
\noindent
Some latest and selected LHC results of searches for additional Higgs bosons using full Run 2 data set are presented.
No strong evidence for the existence of extra Higgs bosons, is found so far.
Several mild excesses were observed by ATLAS and CMS collaborations.
More data is needed to conclude on the nature of these excesses.
\vspace*{0.5cm}

{\bf Acknowledgements.}
\vspace*{0.2cm}

The author would like to thank the CKM2023 organizers for their hospitality and
the wonderful working environment.
The author acknowledge the support from
National Natural Science Foundation of China (12061141003, 11875275) and
China Ministry of Science and Technology (2023YFA1605800, 2022YFE0116900).

\bibliographystyle{amsplain}

\end{document}